\documentclass[pdftex,dvipsnames,usenames,aps,superscriptaddress,twocolumn,amsmath,showkeys]{revtex4-1}

\usepackage{hyperref}
\usepackage{float}
\usepackage{dcolumn}
\usepackage{graphicx}
\usepackage{color}
\usepackage{amssymb}
\usepackage{bold-extra}
\usepackage{caption}
\usepackage{subcaption}

\def\figref#1{Fig.~\ref{fig:#1}}
\def\figlab#1{\label{fig:#1}}  
\def\eqref#1{Eq.~(\ref{eq:#1})}

\def\eqlab#1{\label{eq:#1}}
\def\vB{\bf{v}\times\bf{B}}
\def\vvB{\bf{v}\times\left(\bf{v}\times\bf{B}\right)}
\def\vz{\bf{v}\times\bf{z}}
\def\vvz{\bf{v}\times\left(\bf{v}\times\bf{z}\right)}

\newcommand{\Omit}[1]{}

\renewcommand{\vec}[1]{\mathbf{#1}}

\def\KVI{KVI-Center for Advanced Radiation Technology, University Groningen, P.O. Box 72, 9700 AB Groningen, The Netherlands}
\def\AIVUB{Astrophysical Institute, Vrije Universiteit Brussel, Pleinlaan 2, 1050 Brussels, Belgium}
\def\VUB{Interuniversity Institute for High-Energy, Vrije Universiteit Brussel, Pleinlaan 2, 1050 Brussels, Belgium}
\def\NIKHEF{NIKHEF, Science Park Amsterdam, 1098 XG Amsterdam, The Netherlands}
\def\IMAPP{Department of Astrophysics/IMAPP, Radboud University Nijmegen, P.O. Box 9010,  6500 GL Nijmegen, The Netherlands}
\def\CWI{Center for Mathematics and Computer Science (CWI), PO Box 94079, 1090 GB Amsterdam, The Netherlands}
\def\TUe{Department of Applied Physics, Eindhoven University of Technology (TU/e), PO Box 513, 5600 MB Eindhoven, The Netherlands}
\def\ASTRON{Netherlands Institute of Radio Astronomy (ASTRON),  Postbus 2, 7990 AA Dwingeloo, The Netherlands}
\def\MPIB{Max-Planck-Institut f\"{u}r Radioastronomie, P.O. Box 20 24,  Bonn, Germany}
\def\UCI{Department of Physics and Astronomy, University of California Irvine, Irvine, CA 92697-4575, USA}
\def\DAS{Department of Astrophysical Sciences, Princeton University, Princeton, NJ 08544, USA}
\begin{document}
\def\DPEE{Department of Physics and Electrical Engineering, Linnéuniversitetet, 35195 V\"axj\"o, Sweden}

\title{Thunderstorm electric fields probed by extensive air showers through their polarized radio emission}

\author{T.~N.~G.~Trinh} \email[]{t.n.g.trinh@rug.nl}  \affiliation{\KVI}
\author{O.~Scholten}  \affiliation{\KVI}   \affiliation{\VUB}
\author{A. Bonardi}\affiliation{\IMAPP}
\author{S.~Buitink} \affiliation{\AIVUB} \affiliation{\IMAPP}
\author{A.~Corstanje} \affiliation{\IMAPP}
\author{U.~Ebert} \affiliation{\CWI} \affiliation{\TUe}
\author{J.~E.~Enriquez} \affiliation{\IMAPP}
\author{H.~Falcke}  \affiliation{\IMAPP} \affiliation{\NIKHEF} \affiliation{\ASTRON} \affiliation{\MPIB}
\author{J.~R.~H\"orandel}  \affiliation{\IMAPP} \affiliation{\NIKHEF}
\author{B. M. Hare}\affiliation{\KVI}
\author{P. Mitra} \affiliation{\AIVUB} 
\author{K. Mulrey} \affiliation{\AIVUB}
\author{A.~Nelles}  \affiliation{\IMAPP} \affiliation{\UCI}
\author{J.~P.~Rachen} \affiliation{\IMAPP}
\author{L.~Rossetto}  \affiliation{\IMAPP}
\author{C.~Rutjes} \affiliation{\CWI}
\author{P.~Schellart} \affiliation{\IMAPP}\affiliation{\DAS} 
\author{S.~Thoudam} \affiliation{\IMAPP} \affiliation{\DPEE}
\author{S.~ter Veen} \affiliation{\IMAPP}
\author{T. Winchen}  \affiliation{\AIVUB}

\date{\today}

\begin{abstract}
We observe a large fraction of circular polarization in radio emission from extensive air showers recorded during thunderstorms, much higher than in the emission from air showers measured during fair-weather circumstances. We show that the circular polarization of the air showers measured during thunderstorms can be explained by the change in the direction of the transverse current as a function of altitude induced by atmospheric electric fields. Thus by using the full set of Stokes parameters for these events, we obtain a good characterization of the electric fields in thunderclouds. We also measure a large horizontal component of the electric fields in the two events that we have analysed.
\end{abstract}


\maketitle

\section{Introduction}
Lightning initiation~\cite{Dubinova:2015} and propagation~\cite{Dwyer:2014} are driven by the electric fields in a thunderstorm.
However, performing measurements of these fields is very challenging due to the violent conditions in thunderclouds.
A non-intrusive method to probe thunderstorm electric fields is through a measurement of radio emission from extensive air showers during thunderstorms~\cite{Schellart:2015}. 

When a high-energy cosmic ray strikes the Earth's atmosphere, it generates many secondary particles, a so-called extensive air shower. The dominant contribution to the radio emission from air showers during fair weather (\textit{fair-weather events}) is driven by the geomagnetic field~\cite{Kahn:1966,Scholten:2008}. Electrons and positrons are deflected in opposite directions due to the Lorentz force, which results in a current perpendicular to the shower axis. As the shower develops, this current varies with altitude, thereby producing radio emission. This radiation is linearly polarized in the $\hat{e}_{\vB}$ direction, where $\bf{v}$ is the velocity of the shower front, $\bf{B}$ is the geomagnetic field, and $\hat{e}$ denotes an unit vector. 
In addition, as the shower propagates, a negative charge-excess builds up in the shower front due to the knock-out of electrons from air molecules by the shower particles. The variation of this charge excess gives rise to a secondary contribution to the emission~\cite{Askaryan:1962,Krijn:2010}. The charge-excess emission is also linearly polarized, but radially with respect to the shower axis. For fair-weather events, we observe a small fraction of circular polarization due to the fact that the time-structures of the radio pulses emitted from the charge-excess component and those from the transverse-current component are different~\cite{Scholten:2016}. Since the charge-excess pulses are delayed with respect to the transverse-current pulses and they are polarized in different directions, the polarization of the total pulse rotates from one direction to the other. 
In our analysis, this is seen as circular polarization where the magnitude and handedness depend on the distance and the azimuth position of the observer with respect to the shower axis.

As shown in~\cite{Schellart:2015}, due to the influences of atmospheric electric fields, intensity and linear-polarization footprints of the showers observed during thunderstorms (\textit{thunderstorm events}) are different from those of fair-weather events.
In this paper we show that thunderstorm events have a larger circular polarization component near the shower axis than fair-weather events. 
We demonstrate quantitatively that this can be explained as being due to the variation of the atmospheric electric field with altitude. 
The electric field changes the direction of the transverse current and thus changes the polarization direction of radio emission. 
The signals from the different layers are emitted in sequence when the air shower front, progressing with essentially the light velocity, $c$, passes through. The emitted radio signals travel with a lower velocity than the shower front, $c/n$, where $n$ is the index of refraction.
Thus, near the shower axis, the pulses from the upper layers arrive with a delay with respect to the pulses from the lower layers resulting in a change of the polarization angle over the duration of the pulse, which is seen as circular polarization in the data.
Therefore, the usage of the circular polarization measurements puts strong additional constraints on the structure of the atmospheric electric fields on top of the information obtained by using only the radio intensity.
Since the circular polarization is due to a re-orientation of the transverse current in the shower front the circular polarization does not depend on the azimuthal orientation of the antenna with respect to the shower axis, unlike is the case for the circular polarization of fair weather events.

In this work, we present data on circular polarization seen in the radio emission of a large number of thunderstorm events close to the shower axis as measured with the LOw-Frequency ARay (LOFAR) radio telescope array, see Sec.~\ref{LOFAR}. In Sec.~\ref{modeling}, we present a toy model to explain the cause of circular polarization of air showers measured during thunderstorms. Two reconstructed thunderstorm events are presented in Sec.~\ref{sec-Efield} to show that circular polarization is essential to obtain additional information about the atmospheric electric fields. Conclusions are given in Sec.~\ref{Conclusion}.

\section{LOFAR and data analysis}
\label{LOFAR}
Data for the present analysis were recorded with the Low-Band Antennas (LBAs) in the core of the LOFAR radio telecope~\cite{Haarlem:2013}. Each LBA consists of two dipoles and records in the frequency range of 10 - 90~MHz. These antennas are grouped into circular stations.
The stations are positioned with increasing density towards the center of LOFAR.
The highest density is at the core where 6 such stations are located in a $\sim$320~m diameter region, the so-called `Superterp'. For the purpose of air shower measurements, these antennas are equipped with ring buffers that can store  the raw voltage traces sampled every 5 ns, up to 5 s. 
A trigger is obtained from a particle detector array, LOfar Radboud air shower Array (LORA), from air showers with a primary energy in excess of 2$\times$10$^{16}$ eV~\cite{Thoudam:2014}.

The data are processed in an off-line analysis~\cite{Schellart:2013}. The arrival direction of the air shower is reconstructed based on the arrival times of the radio signals in all antennas. The primary energy of the air shower is estimated by using the particle detector data. 
The radio signal containing the pulse is received by an antenna where the signal-amplitude $S_i$ is determined at 5 ns time intervals, i.e. sampled at 200 MS/s, where the sample number is denoted by $i$.
For each antenna, the Stokes parameters, $I$, $Q$, $U$ and $V$, are expressed as
\begin{equation}
\eqlab{Stokes}
\begin{aligned}
I &= \frac{1}{n}\sum_{i=0}^{n-1}\left(\left |\varepsilon_{i,\vB}  \right |^2+\left |\varepsilon_{i,\vvB}  \right |^2\right),
\\
Q &=  \frac{1}{n}\sum_{i=0}^{n-1}\left(\left |\varepsilon_{i,\vB}  \right |^2-\left |\varepsilon_{i,\vvB}  \right |^2\right),
\\
U +iV&= \frac{2}{n}\sum_{i=0}^{n-1}\left(\varepsilon_{i,\vB}\varepsilon_{i,\vvB}^*\right),
\end{aligned}
\end{equation}
as derived in Ref.~\cite{Schellart:2014}. 
$\varepsilon_i =  S_i+i\hat{S}_i$ are the complex signal voltages, where $\hat{S}_i$ is sample $i$ of the Hilbert transform of $S$. The summation is performed over n = 5 samples, centered around the peak of the pulse.
Stokes $I$ is the intensity of the radio emission. Stokes $Q$ and $U$ are used to derive the linear-polarization angle 
\begin{equation}
\psi = \frac{1}{2}\text{tan}^{-1}\left(\frac{U}{Q}\right)\,,
\label{linear-angle}
\end{equation}
and Stokes $V$ represents the circular polarization.  
\begin{figure*}
\centering
\includegraphics[width=\textwidth]{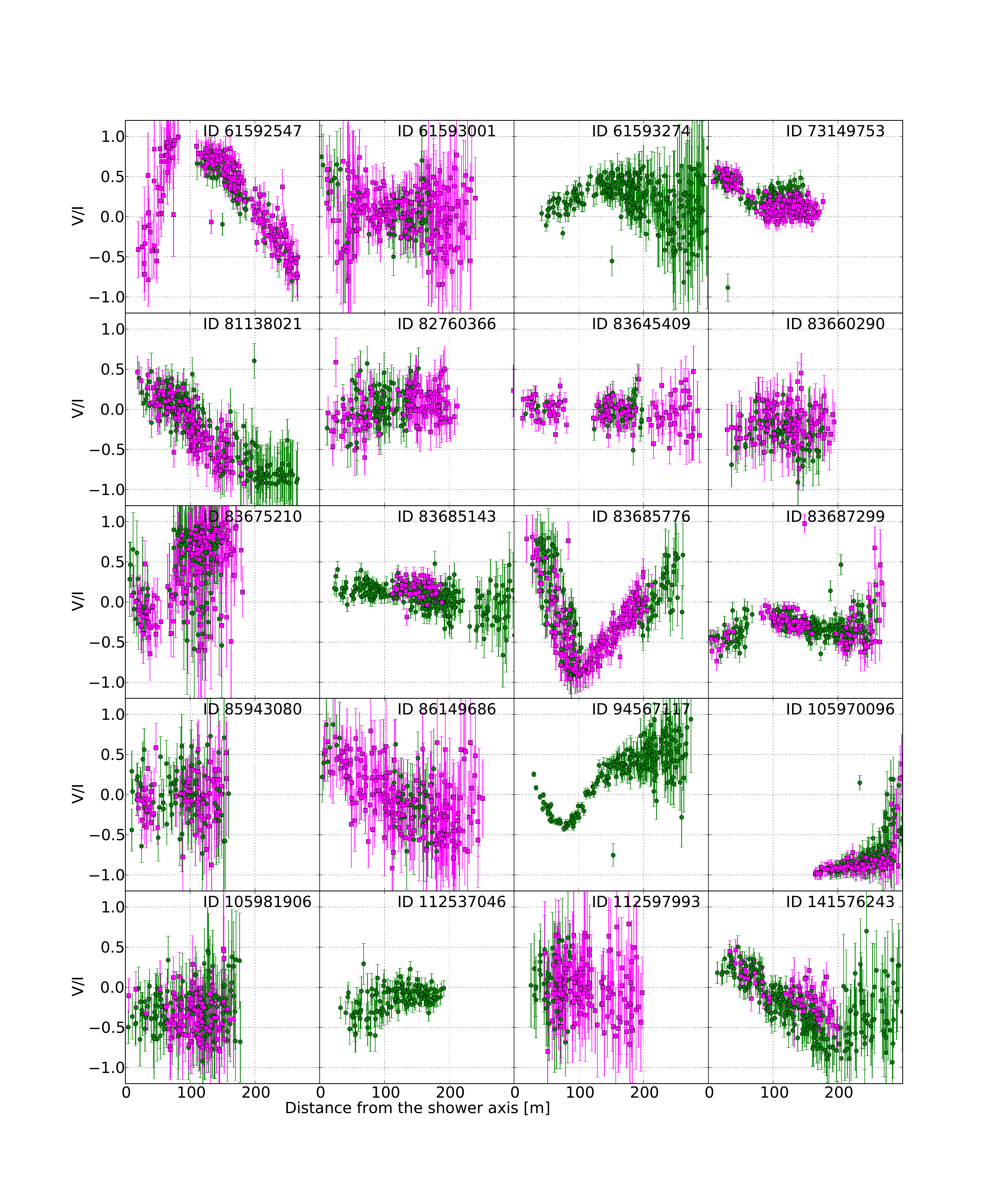}
\caption{The circular polarization for thunderstorm events as a function of distance from the shower axis. Green circles represent the circular polarization at the antennas having an azimuthal position $\varphi = 0^{\circ} - 180^{\circ}$ and purple squares show those for $\varphi = 180^{\circ} - 360^{\circ}$, where $\varphi$ = 0 lies on the positive $\hat{e}_{\vB}$ axis. The ID numbers are used to label the air showers.}
\figlab{coreV_thunderstorm}
\end{figure*}

\begin{figure*}
\centering
\includegraphics[width=\textwidth]{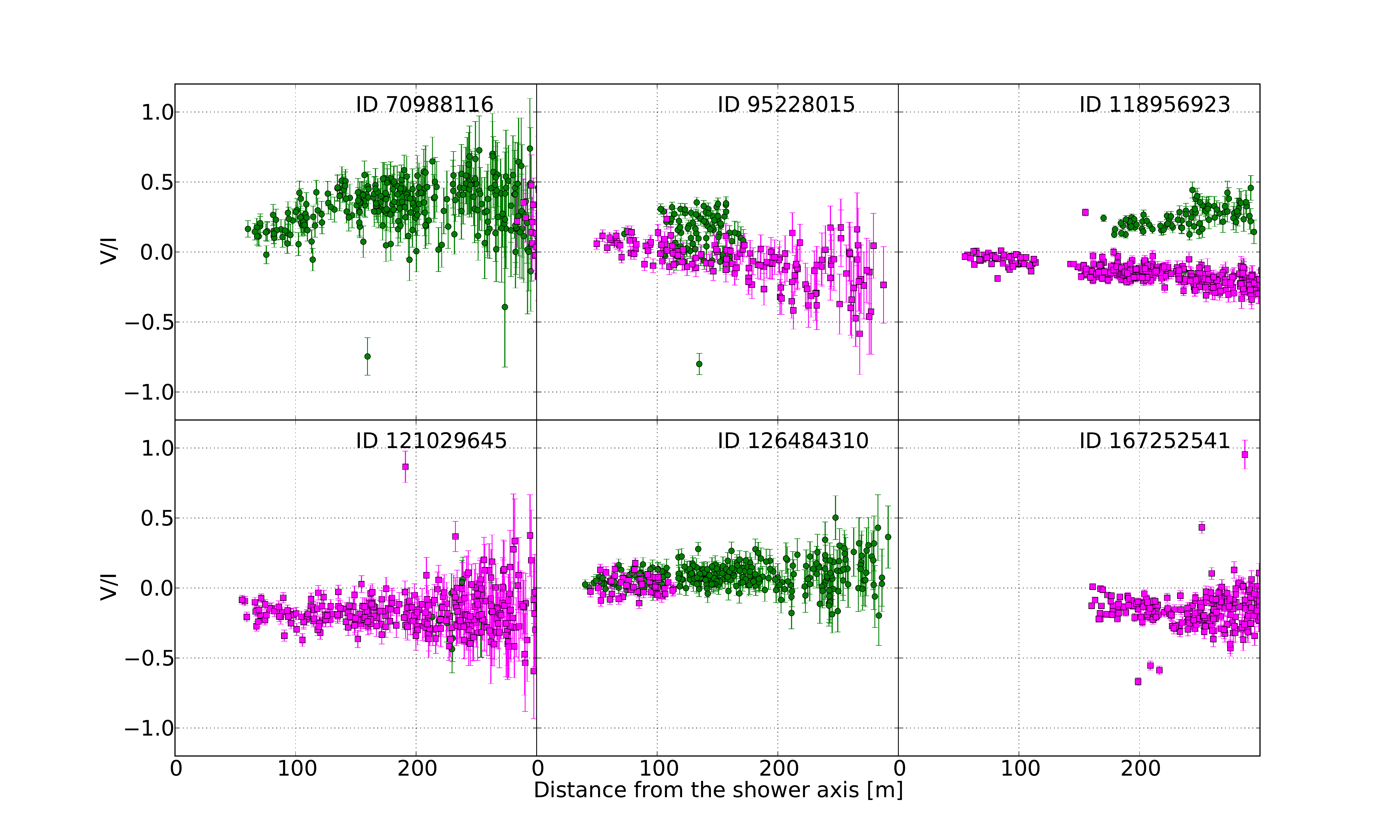}
\caption{The circular polarization for fair-weather events as a function of distance from the shower axis. Green circles represent the circular polarization at the antennas having an azimuthal position $\varphi = 0^{\circ} - 180^{\circ}$ and purple squares show those for $\varphi = 180^{\circ} - 360^{\circ}$, where $\varphi$ = 0 lies on the positive $\hat{e}_{\vB}$ axis. The ID numbers are used to label the air showers.}
\figlab{coreV_fairweather}
\end{figure*}

During the period between June 2011 and January 2015, there were 118 fair-weather events~\cite{Buitink:2016} and 20 thunderstorm events~\cite{Schellart:2015} with radio signals detected in at least 4 LBA stations. 
For comparison, the circular polarization for 20 thunderstorm events and for 6 fair-weather events is shown in~\figref{coreV_thunderstorm} and ~\figref{coreV_fairweather}, respectively. The circular polarization for fair-weather events is very small near the shower axis and increases with distance~\cite{Scholten:2016}. Therefore, in order to show the dependence on azimuth angle, $\varphi$, we seleted those fair-weather events that have data of at least 4 LBA stations beyond 150~m from the shower axis and where the uncertainties in the amount of circular polarization is less than 0.2. 
As can be seen from~\figref{coreV_thunderstorm} and~\figref{coreV_fairweather}, there are significant differences between the circular polarization for the thunderstorm events and that for the fair-weather events.
Firstly, the circular polarization for the thunderstorm events does not depend on the azimuthal position, $\varphi$, of the antenna while for the fair-weather events it is proportional to $\sin\varphi$. Secondly, the circular polarization for some thunderstorm events changes sign at some distances while the dependence of the circular polarization on distance is almost the same for all fair-weather events as mentioned above.
In~\figref{coreV_thunderstorm}, it can be seen that there are some thunderstorm events having very small amount of circular polarization. These events are distinguished from fair-weather events by the linear polarization which has been discussed in Ref.~\cite{Schellart:2015}.
Thirdly, the circular polarization for all fair-weather events is small near the shower axis while it varies from event to event for thunderstorm events. 
This difference is also shown in~\figref{meanV}, where the amount of circular polarization ($|V|/I$) within a 30~m radius of the shower axis is given for 884 antennas recording fair-weather data and 183 antennas taking thunderstorm data.
We choose the radius of 30 meters to concentrate on the near-axis region while also keeping an area large enough to contain a sufficient number of antennas.
The uncertainties indicated in~\figref{meanV} are determined from a Monte Carlo procedure. 
For 500 trials per antenna the Stokes parameters $Q_t$, $U_t$ and $V_t$ are chosen randomly from a Gaussian distribution where the mean and the standard deviation of the distribution correspond to the actual measurement. The Stokes $I_t$ of each trial is calculated by using $I_t^2 = (Q_t^2+U_t^2+V_t^2) + W^2$ where $W$ is calculated from the actual Stokes parameters measured by the antenna, $W^2 = I^2 - \left(Q^2+U^2+V^2\right)$. 
The spread (standard deviation) of the determined distribution of $|V_t|/I_t$ gives the uncertainty.
\figref{meanV} shows that the amount of circular polarization near the shower axis is consistently small for fair-weather events, while a large spread is seen for thunderstorm events.
In Ref.~\cite{Scholten:2016} it was shown that for the fair-weather events the measured circular polarization is well understood. 
The physics of the measured circular polarization of the thunderstorm events is explained in detail in the following section.
\begin{figure}
        \centering
                \includegraphics[width=0.45\textwidth]{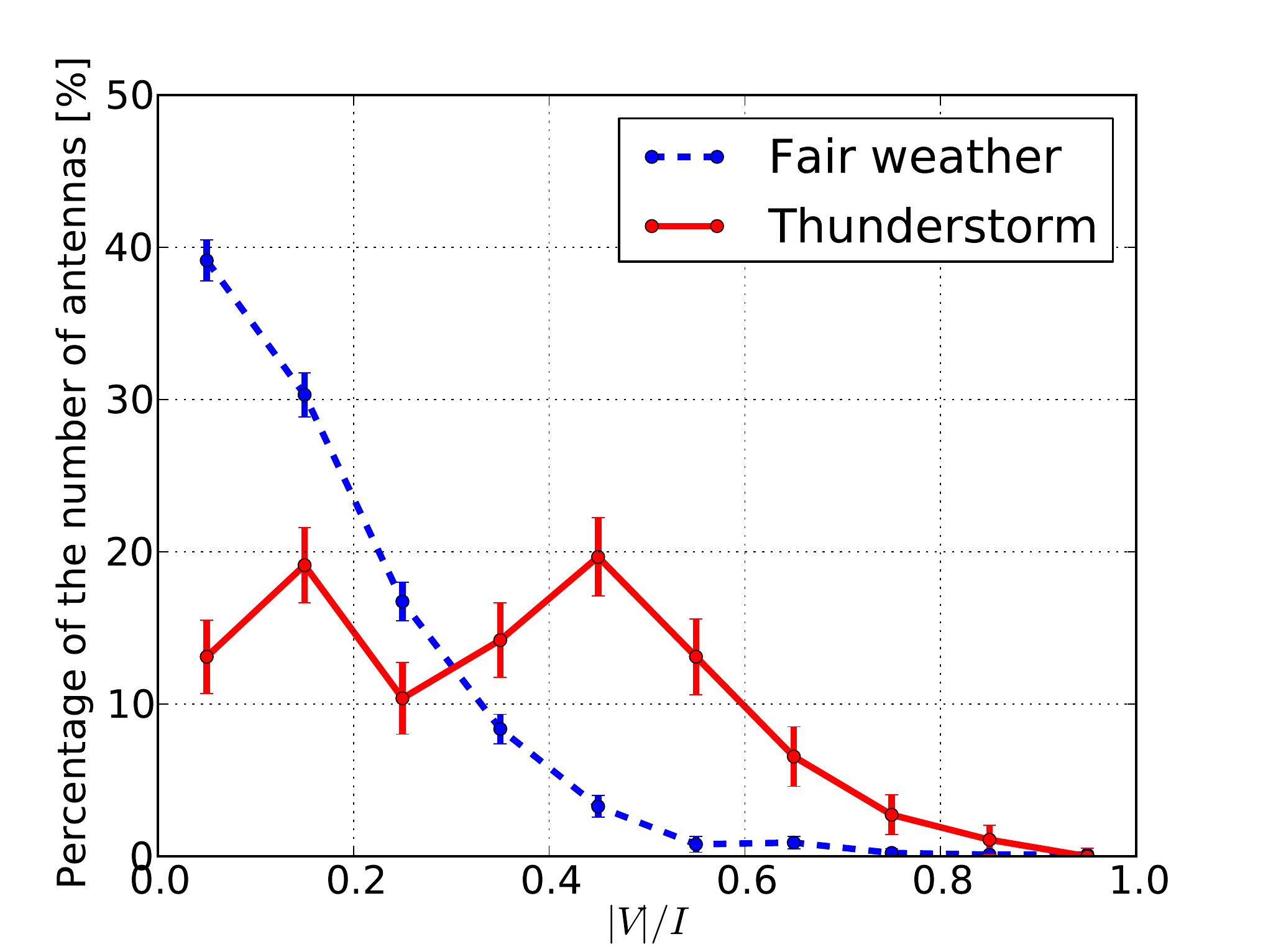}
   \caption{Distribution of the amount of circular polarization in the core of radius 30~m for showers measured during fair weather and thunderstorms.}
   \figlab{meanV}
\end{figure}

\section{Modeling}
\label{modeling}
During thunderstorms the emission of radio waves from air showers is affected by atmospheric electric fields~\cite{Buitink:2007, Apel:2011,Schellart:2015}. The atmospheric electric field can be decomposed into two components $\bf{E}_\perp$ and $\bf{E}_\parallel$, which are perpendicular and parallel to the shower axis, respectively. 
$\bf{E}_\parallel$ increases the number of either electrons or positrons, depending on its orientation, and decreases the other~\cite{Schellart:2015, Trinh:2016}. 
Since the field compensates the energy loss of low-energy electrons, they `live' longer and can thus trail further behind the shower front. As a result, the radiation from these particles does not add coherently in the frequency range 30-80 MHz of the LOFAR LBAs.. 
The transverse component of the field $\bf{E}_\perp$ does not change the number of electrons and positrons, but changes the net transverse force acting on the particles~\cite{Schellart:2015, Trinh:2016}
\begin{equation}
\vec{F}_\perp = q(\vec{E}_{\perp} + \vec{v}\times\vec{B}).
\end{equation}
Hence, the magnitude and the direction of the induced transverse current change according to the net force $\bf{F}_\perp$.
Since for the presented data the influence of the transverse component $\bf{E}_\perp$ on the radio emission dominates, the parallel component $\bf{E}_\parallel$ is set to zero in this work.

The transverse electric field changes the direction of the transverse current, so it also modifies the polarization of the transverse-current radiation. In thunderclouds, not only the magnitude but also the orientation of electric fields changes with altitude~\cite{Marshall:1995}. 
This causes a change of the transverse current in the thunderclouds and thus the linear polarization changes with time.
As explained in the introduction, this results in a changing linear polarization angle over the duration of the pulse giving rise to a large value for V (see \eqref{Stokes}), the component of circular polarization of the pulse.

We use a toy model to show the physics of large circular polarization of the pulses in some of the thunderstorm events. We consider the geometry given in~\figref{observer} as an example. A vertical air shower passes through two layers where the electric field in each is constant. The fields are such that the net forces are perpendicular to each other and make an angle $\varphi$ with $\hat{e}_{\vB}$ as shown in~\figref{observer}. The induced current in the shower front is propotional to the number of particles in the shower multiplied by the net force acting on them. The induced currents thus have orthogonal directions in the two layers where the peak of the current occurs at height $h_m$, corresponding to $X_{\text{max}}$ of the shower, defined as the atmospheric (slant) depth where the number of air-shower particles reaches a maximum. For this case we consider the pulses emitted with a central frequency $\omega$ when the shower passes through each layer
\begin{equation}
\begin{aligned}
\varepsilon_a &= A_ae^{i\left(\omega t+\eta_a\right)}\,,
\\
\varepsilon_b &= A_be^{i\left(\omega t+\eta_b\right)}\,,
\end{aligned}
\end{equation} 
where $\eta =\eta_a-\eta_b=\omega\Delta\,t$ is the phase difference corresponding to an arrival-time difference $\Delta t$ between the two pulses for an observer. 
In thunderstorms, the transverse current is generally enhanced by the atmospheric electric field, so its radiation is much larger than the charge-excess emission and thus we ignore the charge-excess contribution. Therefore, neither $\varepsilon_a$, $\varepsilon_b$ nor $\eta$ depends on the azimuth angle of the antenna position with respect to the shower axis.
Since the transverse currents in the two layers are perpendicular to each their pulses are polarized in two perpendicular orientations on the ground. These pulses can be expressed as
\begin{equation}
\begin{aligned}
\varepsilon_{\vB} &= A_ae^{i\left(\omega t+\eta_a\right)}\cos\varphi - A_be^{i\left(\omega t+\eta_b\right)}\sin\varphi\,,
\\
\varepsilon_{\vvB} &= A_ae^{i\left(\omega t+\eta_a\right)}\sin\varphi + A_be^{i\left(\omega t+\eta_b\right)}\cos\varphi\,.
\end{aligned}
\end{equation} 
Substituting these into \eqref{Stokes} we obtain the Stokes parameters
\begin{equation}
\eqlab{Stokes_ex}
\begin{aligned}
I &= A_a^2+A_b^2\,,
\\
Q &= \left(A_a^2-A_b^2\right)\cos\,2\varphi - 2A_aA_b\sin 2\varphi\cos\eta\,,
\\
U &=  \left(A_a^2-A_b^2\right)\sin\,2\varphi + 2A_aA_b\cos 2\varphi\cos\eta\,,
\\
V &= 2\,A_aA_b\sin\eta\,.
\end{aligned}
\end{equation}

For the special case when $\varphi$ = 0, i.e. the net force in the upper layer is along $\hat{e}_{\vB}$ and the one in the lower layer is along $\hat{e}_{\vvB}$ (see \figref{observer}), the phase shift can be derived from the Stokes $V$ and $U$ parameters
\begin{equation}
\eqlab{eta}
\eta_{\varphi = 0}= \arctan\left(\frac{V}{U}\right)\,.
\end{equation}
\begin{figure}
        \centering
                \includegraphics[width=0.4\textwidth]{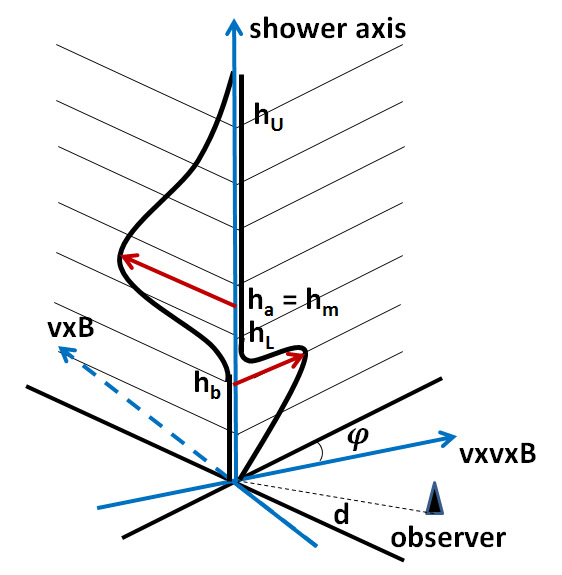}
   \caption{The geometry used in the calculation and a typical current profile of a shower passing through a two-layered electric field where the orientation of the fields in the two layers are perpendicular to each other. The two solid red arrows indicate the net forces acting on air-shower particles.}
   \figlab{observer}
\end{figure}
We will show for this special case how $\eta$ depends on the distance $d$ from the shower axis for a fixed frequency $\omega$. To simplify the calculation, we assume that at the height of $h_a$ the current points in $\hat{e}_{\vB}$ and emits radiation. After that the shower propagates down with the velocity $c$, and the current rotates to $\hat{e}_{\vvB}$ at the height of $h_b=h_a-\Delta\,h$ and radiates another signal. 
The pulses emitted at different heights move with the reduced velocity $v=c/n$ and thus arrive with a time delay due to the fact that index of refraction $n$ is larger than unity.
The signals with a frequency $\omega$ which an observer at a distance $d$ from the origin receives (since $\varphi = 0$) are
\begin{equation}
\begin{aligned}
\varepsilon_{\vB} &= A_a e^{i\omega\left( t-R_a/v\right)}\,,
\\
\varepsilon_{\vvB} &= A_b e^{i\omega\left( t-\Delta\,h/c - R_b/v\right)}\,,
\end{aligned}
\end{equation}
where $R_a=\sqrt{h_a^2+d^2}$ and $R_b=\sqrt{h_b^2+d^2}$ are the distances from the observer to the emision points and $v$ is the velocity of the signals. $\Delta\,h/c$ accounts for the later arrival of the current at $h_b$. The phase shift of these two signals can be derived from \eqref{eta} 
\begin{equation}
\eqlab{eta_a}
\tilde{\eta} =\frac{\omega}{c}\left[n\left(\sqrt{h_b^2+d^2}-\sqrt{h_a^2+d^2}\right)+\Delta\,h\right]\,.
\end{equation}
The phase shift $\tilde{\eta}$ is positive which means that the signal radiated at $h_a$ arrives earlier than the one at $h_b$. For $\tilde{\eta}=0$, the two signals arrive at the observer at the same time. 
Note that~\eqref{eta_a} can only be used in the case where the two emission components are perpendicular to each other and one of the components is along $\hat{e}_{\vB}$. 

For comparison with the analytic calculation, we simulated three vertical showers with CoREAS~\cite{Huege:2013} that included two-layered electric fields with the boundaries between electric fields at different altitudes $h_L$. The electric field EFIELD option~\cite{Buitink:2010} was implemented in CORSIKA~\cite{Heck:1998}. The electric fields in the two layers are such that the net force in the upper layer points in $\hat{e}_{\vB}$ and the one in the lower layer points in $\hat{e}_{\vvB}$, which introduces two perpendicular transverse currents. The upper layer, with strength $|\bf{E}_U|$ = 50~kV/m, starts at a height $h_U$ = 8~km above the ground and extends down to heights of $h_L$ = 4~km, 3~km and 2~km for each simulation. At $h_L$ the lower layer starts and the field strength decreases to $|\bf{E}_L|$  = 25~kV/m. The shower maximum $X_{\text{max}}$ = 580 g/cm$^2$ is the same in all three simulations, corresponding to $h_m\approx$ 4.6~km which is in the upper layer. 

In order to be compared with the analytic calculation where pulses are assumed to emit a central frequency $\omega$, the phase shift $\eta_C$ from the CoREAS simulations in the narrow frequency band, $60-65$~MHz, is derived and displayed in~\figref{eta}.
The phase shift $\tilde{\eta}$ derived from \eqref{eta_a} is also shown in~\figref{eta} for $\omega$ = 65~MHz. To simplify the calculation, the refractive index is kept constant at $n=1.00015$. Note that the heights $h_a$ and $h_b$ in \eqref{eta_a} are the average heights from which the dominant intensity is emitted for the two polarization directions and are thus not equal to $h_U$ and $h_L$. In the upper layer, the maximum emission occurs at $h_a=h_m$. In the lower layer, the height $h_b$ depends on the distance from the observer to the shower axis. At large distances, beyond $\sim$~50~m, the maximum emission arrives from ${h}'_b = h_L - X_a/\rho$, where the air density $\rho$ is approximately~\cite{Marshall:1995} $\rho(h) = 1.208\cdot 10^{-3} \exp(-h/8.4)\;\text{g/cm$^3$}$ and $X_a$ is the adapting distance varying with heights (see Fig. 20 in Ref.~\cite{Trinh:2016}). The average values of $h_b$ for the three simulations are 3.2~km, 2.53~km and 1.78~km, respectively. At the distance $d < {h}'_b\tan\theta={h}'_b\sqrt{n^2-1}$, where $\theta$ is the opening angle corresponding to the distance $d$, the observer receives the dominant signal from ${h}'_b = d/\sqrt{n^2-1}$. As seen in~\figref{eta}, it is the distance at which all three lines coincide. 
At large distances, $\tilde{\eta}$ is positive which means the observer receives the signal radiated at $h_a$ first and the one at $h_b$ later.
At about 50~m, the two signals arrive at the observer at the same time. 
It can be seen from~\figref{eta} that the calculation agrees quite well with the simulations, which demonstrates that the source of the circular polarization is well-understood.
However, for more general geometries of atmospheric electric fields, the layer heights, field strengths and field orientations can only be found through a numerical optimization procedure.
\begin{figure}
        \centering
                \includegraphics[width=0.5\textwidth]{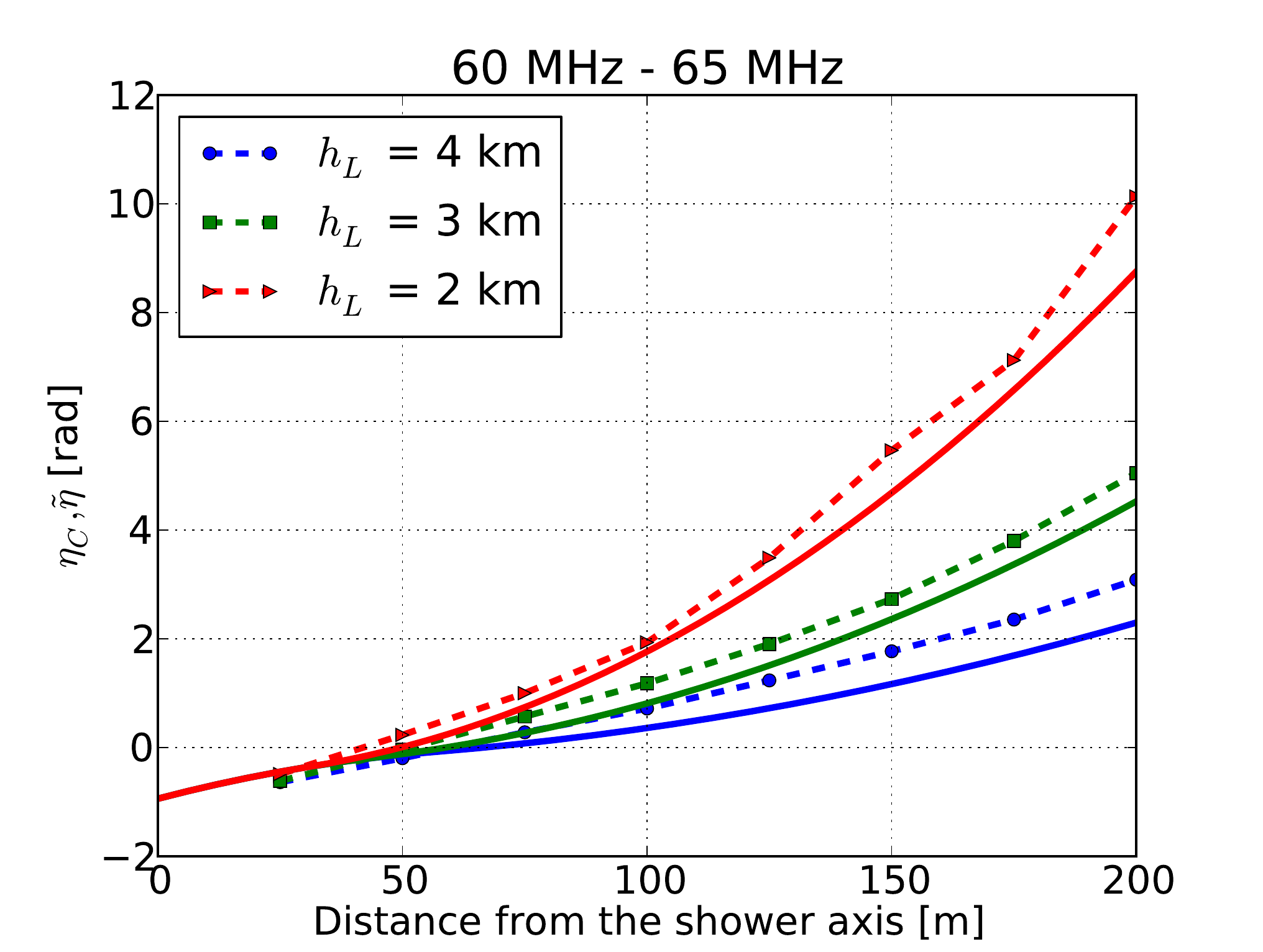}
   \caption{The phase shift $\eta$ as a function of distance from the shower axis. Dotted curve: phase shift $\eta_C$ from CoREAS results. Solid curves: phase shift $\tilde{\eta}$ from an analytic calculation.}
   \figlab{eta}
\end{figure}

\section{Probing the structures of atmospheric electric fields}
\label{sec-Efield}
As discussed in the previous section, the circular polarization in thunderstorm events is caused by the variation in the orientation of the atmospheric electric fields. 
Therefore, using the full set of Stokes parameters, i.e. the combination of intensity, linear polarization, and circular polarization, will allow a more accurate determination of the electric fields in the cloud-layers where the air shower passes through than when using only intensity information as in Ref.~\cite{Schellart:2015}.
To provide more insight into this assertion, we discuss in detail the rescontruction of two thunderstorm events which are called in this work event No.1 and event No.2.

Fitting thunderstorm events is challenging since the electric fields contain many parameters. Another problem is that since CoREAS is a Monte-Carlo simulation, two calculations with similar electric fields can give considerably different results due to shower-to-shower fluctuations even when using the same random seed. Therefore, to determine the electric fields, we first perform a fit using a semi-analytic calculation~\cite{Scholten:2016a} of the radio footprint of air showers based on the current profile. This procedure requires much less CPU time and there are no shower to shower fluctuations. This allows for a standard steepest descent fitting procedure. 
Since this method only approximates the structure of the shower front, we use this to get close to the optimal choice after which we use CoREAS for the final calculations. In order to obtain a prediction of the two-dimensional footprints of the four Stokes parameters, we run CoREAS simulations for 160 antennas which form  a star-shaped pattern with eight arms as in Ref.~\cite{Buitink:2013}, and make an interpolation to reconstruct the full profile. The results are filtered in the frequency range of the LOFAR LBAs. 

The electric field fields are labeled with indices 1, 2 and 3 where 1 is the top layer. 
Each layer is defined by the height $h$ above the ground where the electric field starts and the field $\vec{E}_\perp$.
Note that our analysis cannot determine the parallel components of the electric fields $\bf{E}_\parallel$, therefore we will always work in the 2D plane perpendicular to $\hat{e}_{\bf{v}}$. In this plane, the perpendicular components, $\bf{E}_\perp$, are expressed in two bases. ${1)}$ It can be expressed as the field strength $|\bf{E}_\perp|$ and the angle $\alpha$ between the net force and $\hat{e}_{\bf{v}\times\bf{B}}$, where the net force is the vectorial sum of the Lorentz force and the electric force given by the electric field. ${2)}$ It can also be decomposed into $E_{\vz}$ and $E_{\bf{v}\times(\vz)}$, the components of $\bf{E}_\perp$ along $\hat{e}_{\vz}$ and  $\hat{e}_{\bf{v}\times(\vz)}$, respectively. Here $\hat{e}_{\bf{z}}$ is vertically pointing up.

The intensity footprint, the linear polarization footprint and the circular polarization footprint  of  thunderstorm event No.1, measured at 12:38:37~UTC, December 30$^{th}$, 2012, are displayed in~\figref{94567117}.
The fractions of Stokes parameters are shown in~\figref{Stokes_94567117}. 
The intensity footprint (top panel of~\figref{94567117}) of this event shows a bean shape which is also observed in fair-weather events.
The differences are that the maximum intensity is not in the $\vB$-direction as it is in fair-weather events and the linear polarization (middle panel~\figref{94567117}) is not oriented mainly along $\hat{e}_{\bf{v}\times\bf{B}}$ as it is in the fair-weather events. The polarization footprint shows a `wavy' pattern near the shower axis where the polarization is different from the one at the outer antennas. 
We observe a large fraction of circular polarization in this event, varying as a function of the distance from the antenna to the shower axis.  This can be seen in the bottom panel of~\figref{94567117} and the right panel of~\figref{Stokes_94567117}. 
Therefore, for this event, using only the intensity footprint gives incomplete information about the atmospheric electric field.
The simplest structure of the electric field which can capture the main features of this event is a three-layer field. 
The reconstruction is optimal for the values of the parameters given in Table~\ref{field1}. 
The simulation has values of $X_{\text{max}}$ = 665~g/cm$^2$. The primary energy of the shower is E = $4.7\times10^{16}$~eV and the zenith angle is $\theta$ = 15.5$^\circ$.
Since we do not observe a ring-like intensity pattern, the emission from different layers should not interfere destructively, and thus the fields should not have opposite orientation as taken in Ref.~\cite{Schellart:2015}.
The change in the orientation of the electric field between the second layer and the third layer, close to the ground, results in a change in the direction of the transverse current and thus gives rise to the rotation of the linear polarization as well as a large amount of circular polarization in the region close to the shower axis. 
Near the shower axis the radio signal is most sensitive to the later stages of the shower development, while at large distances the currents higher in the atmosphere have more weight. Thus, a much smaller circular polarization component is observed at larger distances.
There are some differences between the measured and simulated Stokes parameters seen in~\figref{Stokes_94567117} since the three-layered electric field is still an oversimplification of the realistic field. 
The reduced $\chi^2$ for a joint fit of both the Stokes parameters and the particle data $\chi^2/\text{ndf}$ = 4.5, which is large compared to $\chi^2/\text{ndf}\approx$ 1 found in fair-weather showers.
However, all main features are captured.
\begin{figure}
\centering
 \begin{subfigure}{0.38\textwidth}
 	\includegraphics[width=\textwidth]{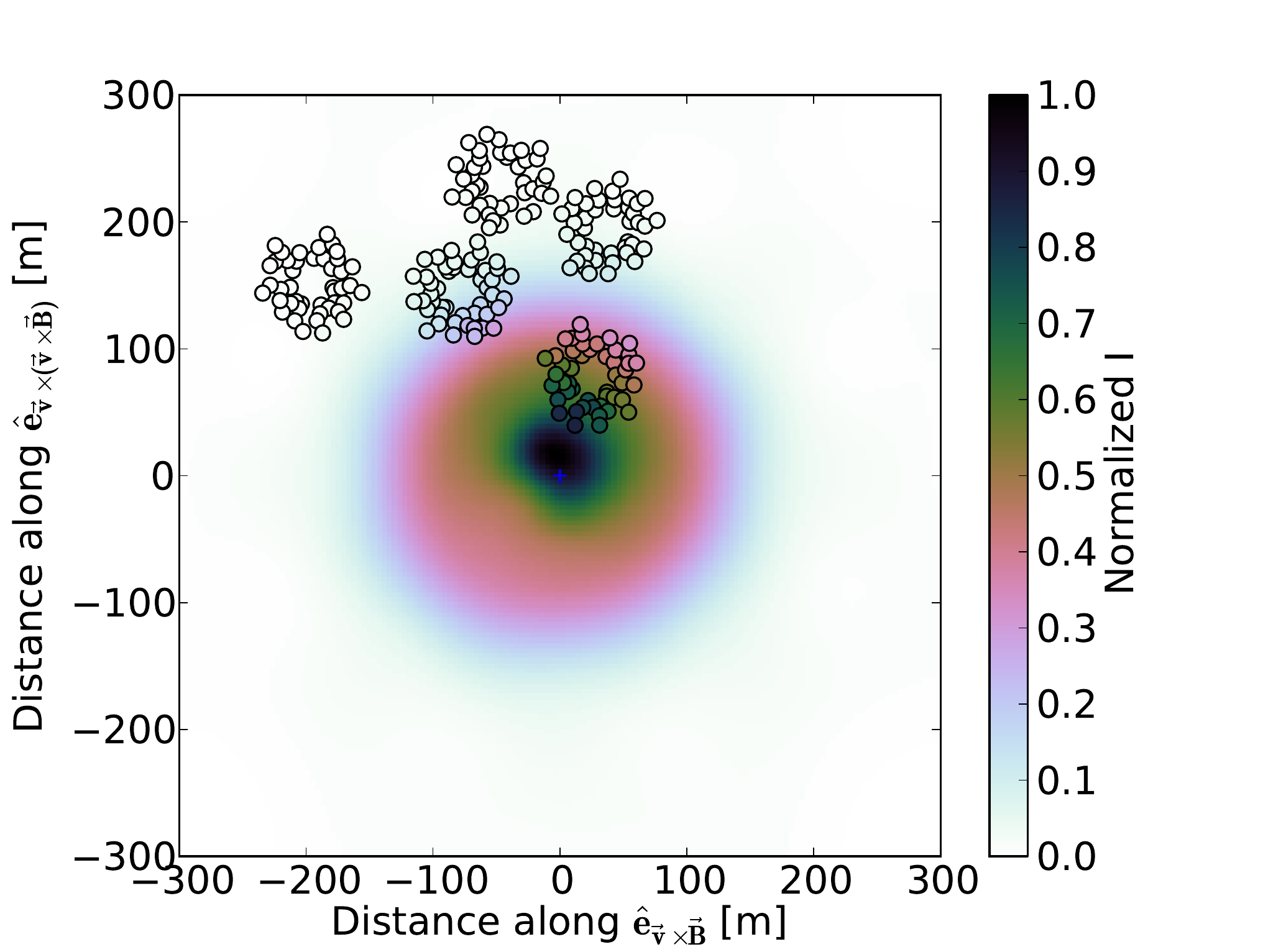}
 	\caption{The intensity (Stokes I parameter) footprint of the thunderstorm event No.1. The background color shows the simulated results while the coloring in the small circles represents the data.}
 \end{subfigure}
\begin{subfigure}{0.35\textwidth}
	\includegraphics[width=\textwidth]{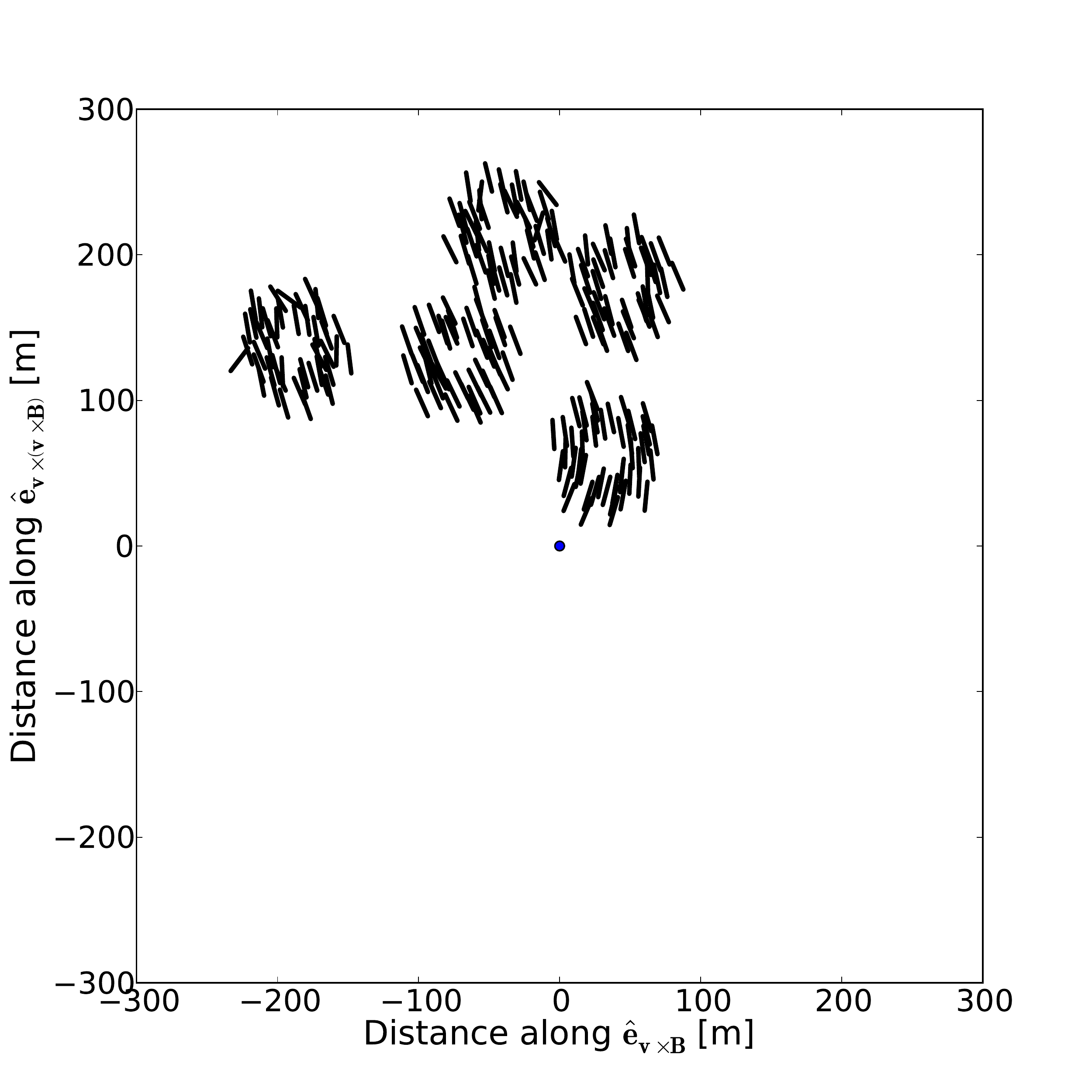}
 	\caption{Linear polarization as measured with individual LOFAR LBAs (lines) in the shower plane.}
  \end{subfigure}
 \begin{subfigure}{0.38\textwidth}
 	\includegraphics[width=\textwidth]{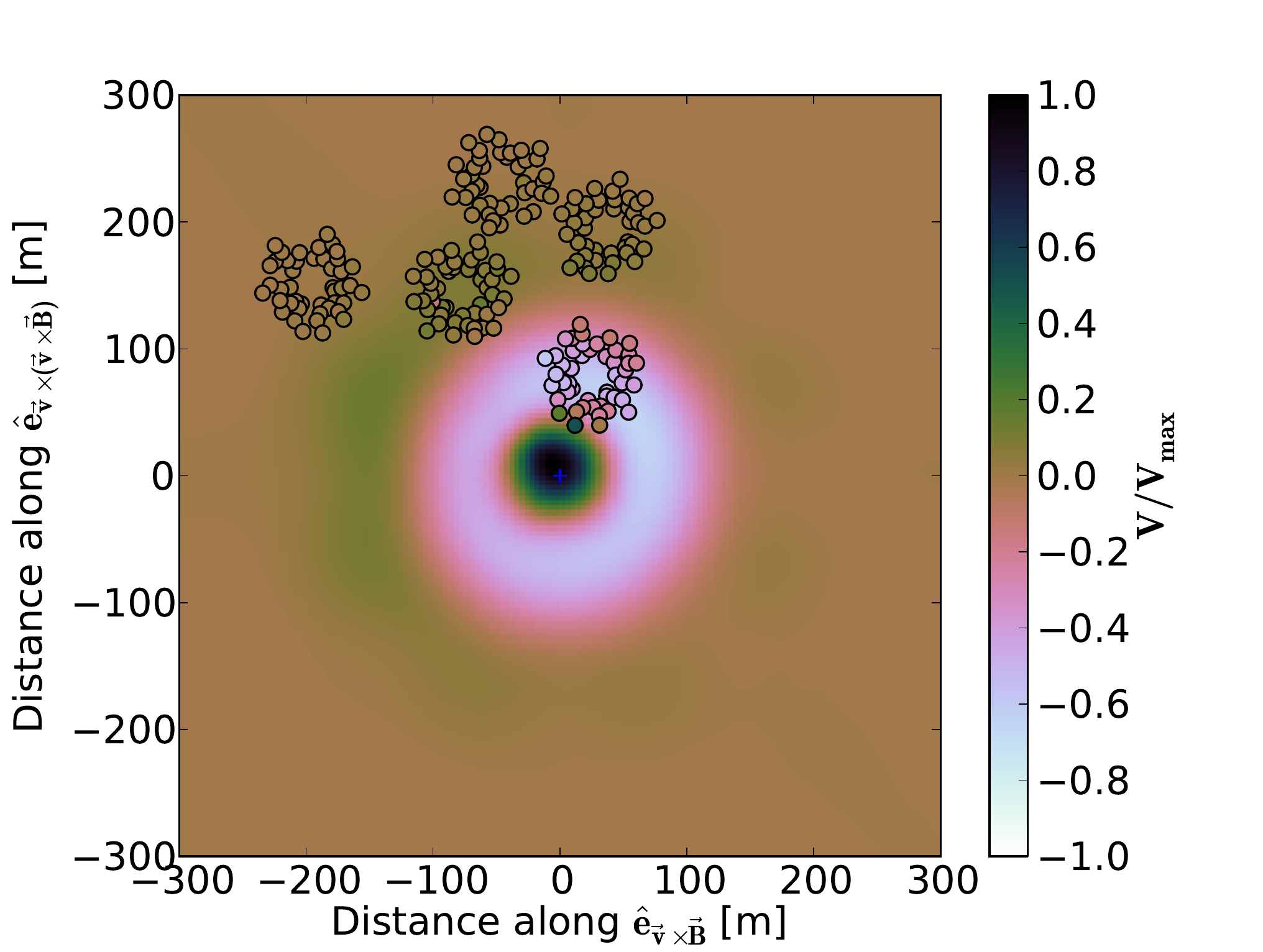}
 	\caption{The footprint of Stokes V parameter, representing the circular polarization. The background color shows the simulated results while the coloring in the small circles represents the data.}
 \end{subfigure}
\caption{Radio polarization footprints of the thunderstorm event No.1. }
\figlab{94567117}
\end{figure}
\begin{figure*}
\centering
\includegraphics[width=1.0\textwidth]{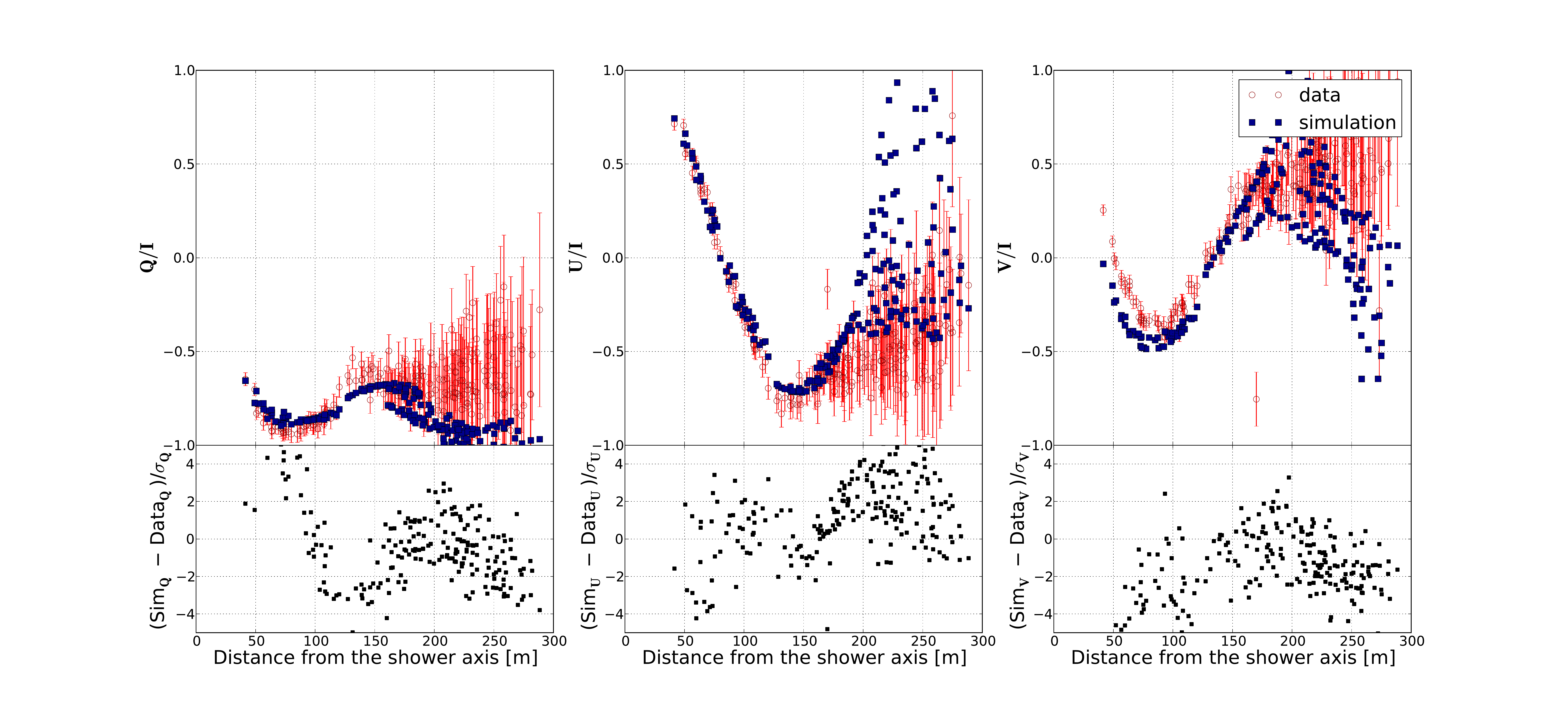}
\caption{The set of normalized Stokes parameters of the thunderstorm event No.1 as recorded with the LOFAR LBAs (open red circles) is compared to the results of the CoREAS simulation (filled blue dots). $\sigma$ denotes one standard deviation error.}
\figlab{Stokes_94567117}
\end{figure*}

\begin{table}
\begin{center}
    \begin{tabular}{ |l | c| c | c|}
    \hline
    Layer & 1 & 2 & 3 \\ \hline
    Height (km) & 8 - 5& 5 - 2 & 2 - 0 \\ \hline
    $|\vec{E}_\perp|$ (kV/m) & 50 & 15 & 9  \\ \hline
    $\alpha$ ($^\circ$) & 98 & 98 & 8 \\ \hline
   $E_{\vz}$ (kV/m) & 46 & 13 & 4 \\ \hline
   $E_{\vvz}$ (kV/m) & -22 & -9 & 8 \\
\hline
    \end{tabular}
\end{center}
 \caption{The structure of the three-layered electric field of the thunderstorm event No.1.}
   \label{field1}
\end{table}

\figref{83687299} shows the intensity footprint and the polarization footprint  of thunderstorm event No.2, which was also presented in Ref. \cite{Schellart:2015}. 
The fractions of Stokes parameters are shown in~\figref{Stokes_83697299}. 
\begin{figure}
\centering
 \begin{subfigure}{0.38\textwidth}
 	\includegraphics[width=\textwidth]{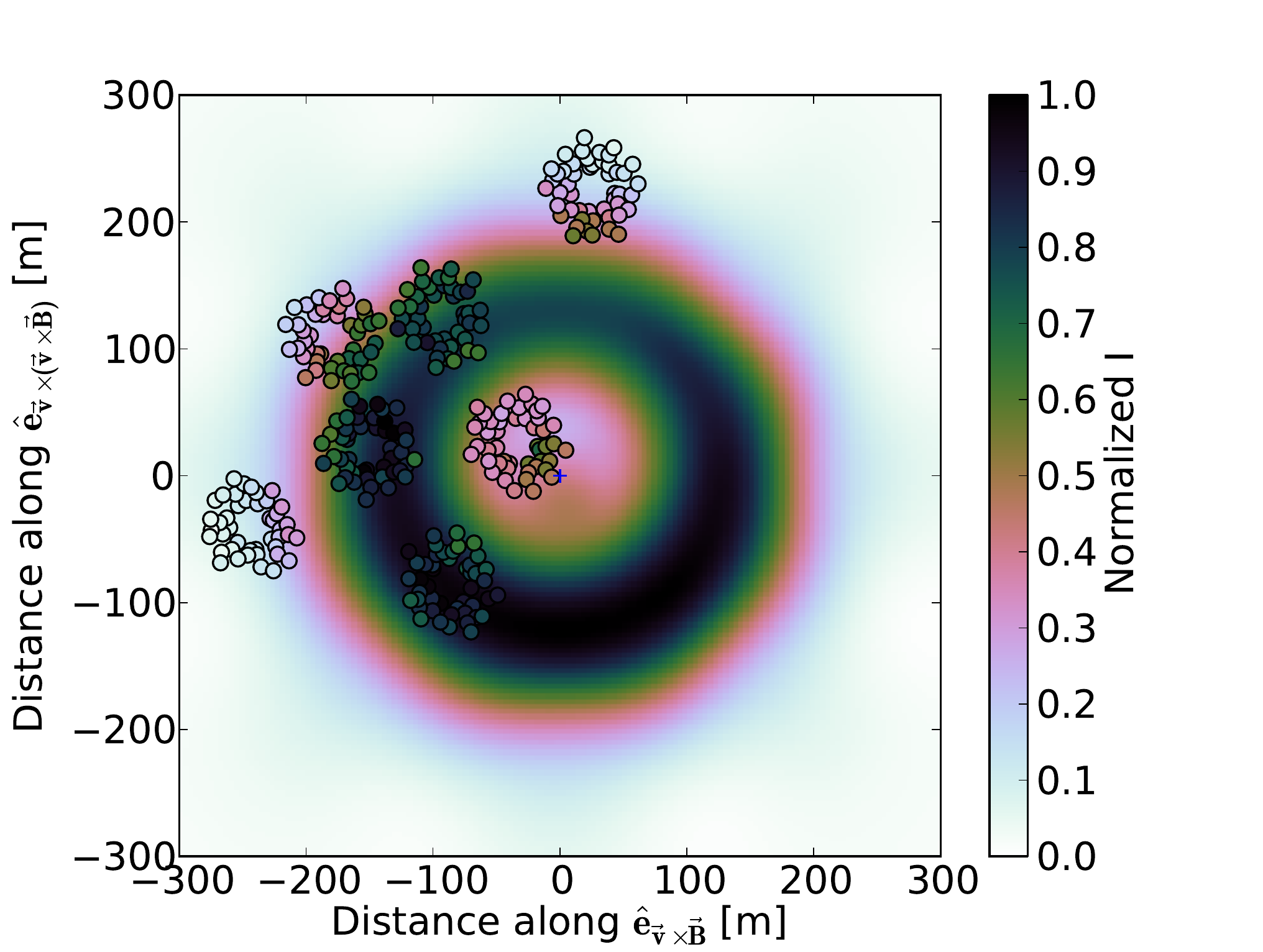}
 	\caption{The intensity (Stokes I parameter) footprint of the thunderstorm event No.2. The background color shows the simulated results while the coloring in the small circles represents the data.}
 \end{subfigure}
\begin{subfigure}{0.35\textwidth}
	\includegraphics[width=\textwidth]{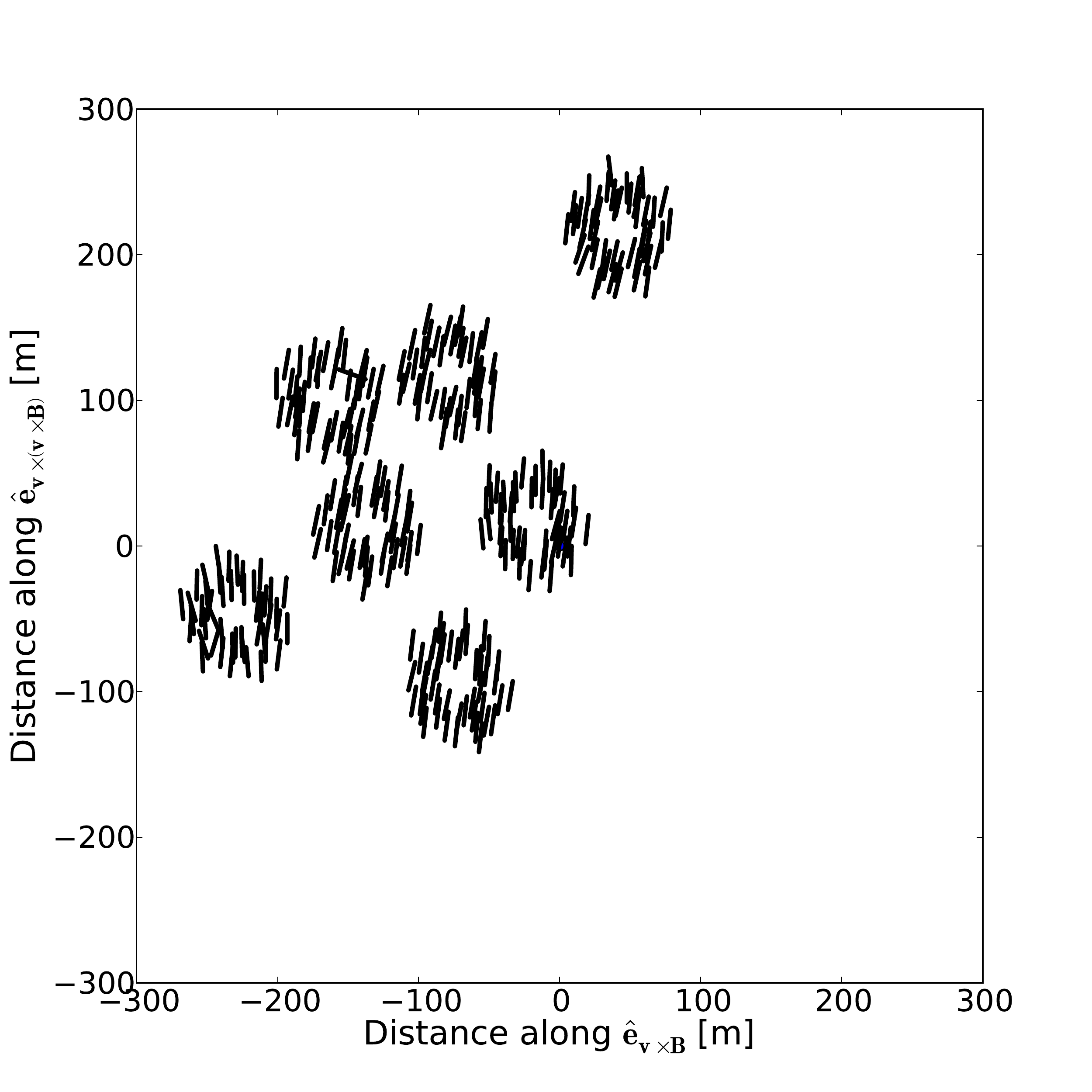}
 	\caption{Linear polarization as measured with individual LOFAR LBA (lines) in the shower plane.}
  \end{subfigure}

 \begin{subfigure}{0.38\textwidth}
 	\includegraphics[width=\textwidth]{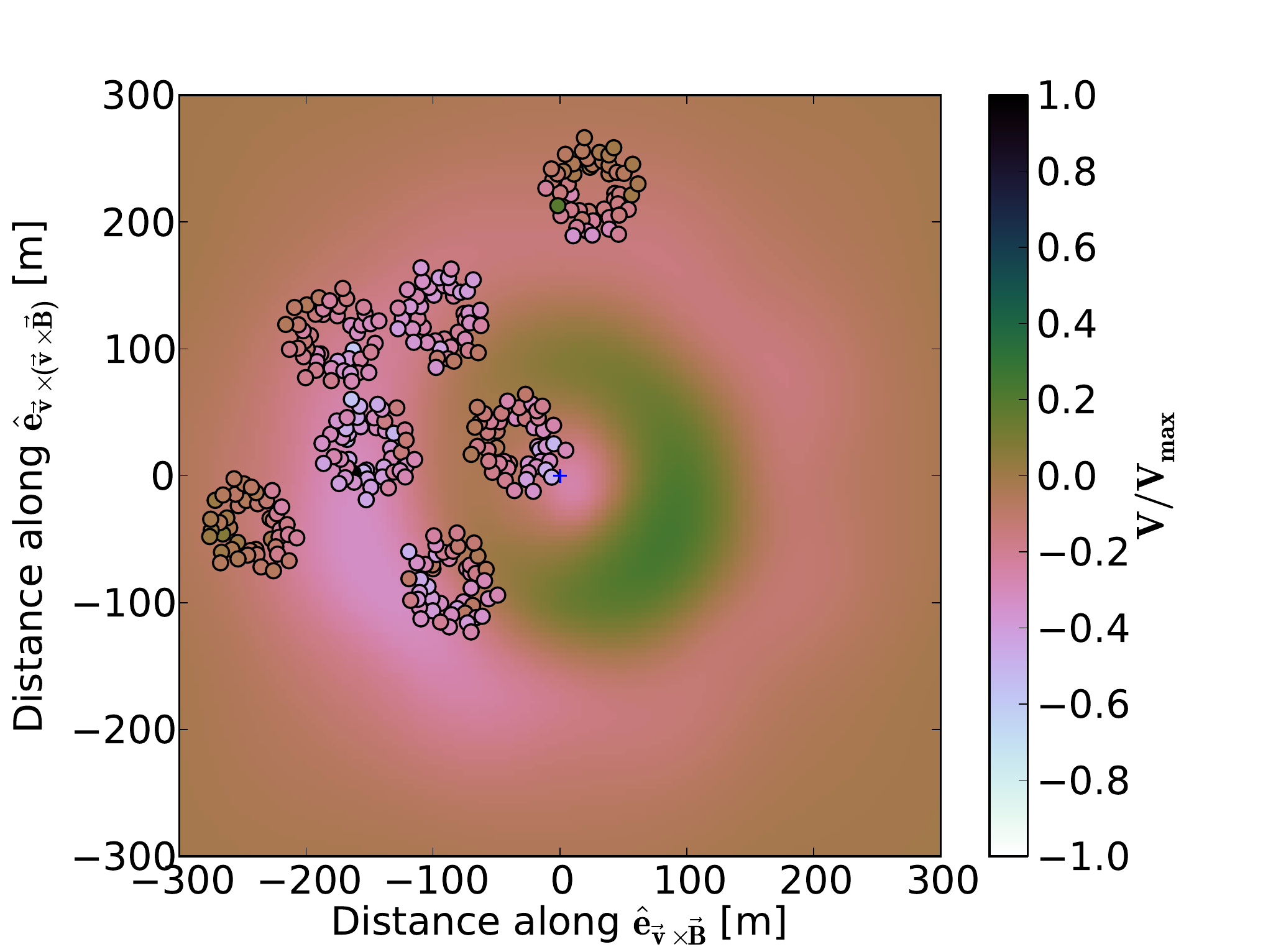}
 	\caption{The footprint of Stokes V parameter, representing the circular polarization. The background color shows the simulated results while the coloring in the small circles represents the data.}
 \end{subfigure}
\caption{Radio polarization footprints of the thunderstorm event No.2. }
\figlab{83687299}
\end{figure}
\begin{figure*}
\centering
\includegraphics[width=1.1\textwidth]{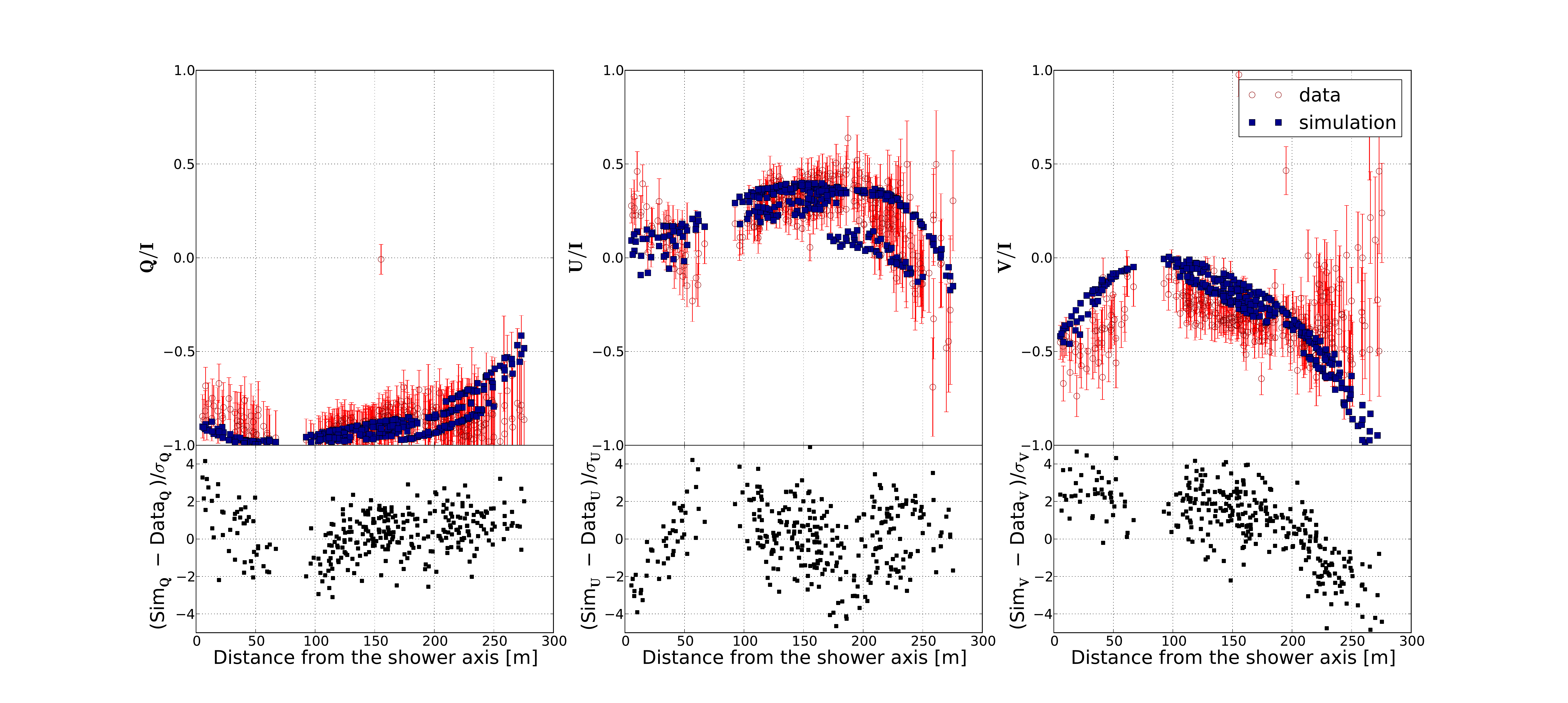}
\caption{The set of normalized Stokes parameters of the thunderstorm event No.2 as recorded with the LOFAR LBAs (open red circles) is compared to the results of the CoREAS simulation (filled blue dots). $\sigma$ denotes one standard deviation error.}
\figlab{Stokes_83697299}
\end{figure*}
This event was measured at 14:28:19~UTC, August 26$^{th}$, 2012.
The ring-like structure in the intensity footprint (top panel of~\figref{83687299}) and the overall polarization direction (middle panel of~\figref{83687299}) indicate that at least a two-layered electric field is needed~\cite{Schellart:2015}, where the electric fields are pointing in opposite directions to introduce a destructive interference between the radiation from the two layers.  
However, the large amount of circular polarization near the shower axis (see the bottom panel of~\figref{83687299} and the right panel of~\figref{Stokes_83697299}) cannot be reproduced by such a field configuration since there is no rotation of the current. The simplest structure of an electric field which can capture the main features of this event is a three-layered field. 
Table.~\ref{field2} presents the values of the electric field giving the best reconstruction of this shower. 
The electric fields obtained here follow the same general structure as presented in our earlier work~\cite{Schellart:2015}. Like in Ref.~\cite{Schellart:2015} the strength of the fields in the lower layer are about half the strength as in the upper layer with almost opposite orientation. However, in the present, more detailed, analysis an additional layer needed to be introduced which shows that the method used in this work gives more accurate information about the electric fields in thunderstorms.
The shower maximum is $X_{\text{max}}$ = 628~g/cm$^2$.
The primary energy of this shower is E = $3.1\times10^{16}$~eV and the zenith angle is $\theta$ = 24.8$^\circ$. 
\begin{table}
\begin{center}
    \begin{tabular}{ |l | c| c | c|}
    \hline
    Layer & 1 & 2 & 3 \\ \hline
    Height (km) & 8 - 6.9& 6.9 - 2.7 & 2.7 - 0 \\ \hline
    $|\vec{E}_\perp|$ (kV/m) & 50 & 20 & 18  \\ \hline
    $\alpha$ ($^\circ$) & -78 & -104 & 67 \\ \hline
   $E_{\vz}$ (kV/m) & -46 & -12 & 14 \\ \hline
   $E_{\vvz}$ (kV/m) & -16 & -16 & 11 \\
    \hline
    \end{tabular}
\end{center}
\caption{The structure of the three-layered electric field of the thunderstorm event No.2.}
   \label{field2}
\end{table}
There is also an almost complete reversal of the electric field from the second layer to the third layer which gives rise to the ring-like structure in the intensity footprint and keeps the linear polarization unique.
The reduced $\chi^2$ for a joint fit of both the Stokes parameters and the particle data is $\chi^2/\text{ndf}$ = 3.5, which is large but reproduces all the main features.

We have checked that the fit quality is sensitive to the heights of the layers on the order of hundred meters and the orientations of the electric fields at the level of degrees. However, it is not sensitive to heights above 8~km because at that height there are few particles in the shower and thus their contribution to the total radio emission is small.
The electric fields shown in Table~\ref{field1} and Table~\ref{field2} only include the components of the true fields perpendicular to the shower axis. The parallel component of the electric fields hardly affects the LBA observations and thus it cannot be determined. In addition, in the frequency domain of the LBAs there is no sensitivity to the component of electric fields in excess of about 50~kV/m, so the strength of the perpendicular component can only be probed up to about this strength (see Ref.~\cite{Trinh:2016} for the discussion). To increase the sensitivity, we would need lower-frequency antennas.

However, as explained in the following we have measured large horizontal components of the electric fields along the shower axis in thunderclouds. 
A strict vertical electric field can be decomposed into two components, one along $\hat{e}_{\bf{v}}$ and the other one along $\hat{e}_{\vvz}$.
Measuring a component in $\hat{e}_{\vvz}$ (see Table~\ref{field1} and Table~\ref{field2}) could thus be a reflection of a vertical field since the present observations have little sensitivity to an $\hat{e}_{\bf{v}}$ component of the electric field.
However, a non-zero component in the $\hat{e}_{\vz}$ direction (see Table~\ref{field1} and Table~\ref{field2}) can never be a projection of a purely vertical electric field, and is thus a genuine signature of a horizontal component.
We have confirmed that setting any of the $E_{\vz}$ components to zero results in poorly reconstructed Stokes parameters. Therefore, it can be concluded that the atmospheric electric field is not fully vertical, but has a significant horizontal component. 
A three-layered structure and a horizontal component of the electric fields in thunderclouds have also been observed in balloon experiments~\cite{Byrne:1983,Marshall:1995, Stolzenburg:2010,Dwyer:2014}.
The large component of a horizontal electric field at high altitudes can be given by two oppositely charged regions inside a thundercloud. The small horizontal component at low altitudes can be given by the main negative-charge layer of a thundercloud in the center and a local positive-charge region at the bottom of the cloud.

\section{Conclusion}
\label{Conclusion}
Air showers measured with the LOFAR LBAs during thunderstorms have generally a much stronger circular polarization component near the shower axis than showers recorded during fair weather. We demonstrate on the bases of a simple model that this is a reflection of the fact that the orientation of atmospheric electric fields changes with height.
This gives rise to a rotation in the direction of the transverse current as the air shower proceeds towards the surface of the Earth. 
This is also confirmed by CoREAS simulations.

Using the full set of the Stokes parameters  thus strongly improves the determination of the atmospheric electric fields in thunderclouds.
As specific examples we have analyzed two thunderstorm events where we show that the intensity and polarization signature can only be described by a three-layered electric field. Also in baloon measurements, generally three different layers are observed below a height of 8~km. 
In our analysis, we also determine that the atmospheric electric field has a sizable horizontal component.  

\begin{acknowledgements}
The LOFAR cosmic ray key science project acknowledges funding from an Advanced Grant of the European Research Council (FP/2007-2013) / ERC Grant Agreement n. 227610. The project has also received funding from the European Research Council (ERC) under the European Union's Horizon 2020 research and innovation programme (grant agreement No 640130).   We furthermore acknowledge financial support from FOM, (FOM-project 12PR304) and NWO (VENI grant 639-041-130). AN is supported by the DFG (research fellowship NE 2031/1-1).

LOFAR, the Low Frequency Array designed and constructed by ASTRON, has facilities in several countries, that are owned by various parties (each with their own funding sources), and that are collectively operated by the International LOFAR Telescope foundation under a joint scientific policy.
\end{acknowledgements}
\bibliography{circular_polar_thunderstorm}
\end{document}